\documentclass[twocolumn,showpacs,preprintnumbers,amsmath,amssymb,superscriptaddress]{revtex4}

\usepackage{graphicx}
\usepackage{dcolumn}
\usepackage{bm}
\begin{document}


\title{Tightly-bound Cooper pair, quasiparticle kinks and clues on the pairing potential in a high $T_c$ FeAs Superconductor}


\author{L. Wray}
\affiliation{Joseph Henry Laboratories of Physics, Department of Physics, Princeton
University, Princeton, NJ 08544, USA}
\author{D. Qian}
\affiliation{Joseph Henry Laboratories of Physics, Department of Physics, Princeton
University, Princeton, NJ 08544, USA}
\author{D. Hsieh}
\affiliation{Joseph Henry Laboratories of Physics, Department of Physics, Princeton
University, Princeton, NJ 08544, USA}
\author{Y. Xia}
\affiliation{Joseph Henry Laboratories of Physics, Department of Physics, Princeton
University, Princeton, NJ 08544, USA}
\author{L. Li}
\affiliation{Joseph Henry Laboratories of Physics,  Department of Physics, Princeton
University, Princeton, NJ 08544, USA}
\author{J.G. Checkelsky}
\affiliation{Joseph Henry Laboratories of Physics,  Department of Physics, Princeton
University, Princeton, NJ 08544, USA}
\author{A. Pasupathy}
\affiliation{Joseph Henry Laboratories of Physics,  Department of Physics, Princeton
University, Princeton, NJ 08544, USA}
\author{K.K. Gomes}
\affiliation{Joseph Henry Laboratories of Physics,  Department of Physics, Princeton
University, Princeton, NJ 08544, USA}
\author{A.V. Fedorov}
\affiliation{Lawrence Berkeley National Laboratory, Advanced Light
Source, Berkeley, CA 94305,
USA}
\author{G.F. Chen}
\affiliation{Beijing National Laboratory for Condensed Matter Physics, Institute of Physics, Chinese Academy of Sciences, Beijing 100080, P.R. China}
\author{J.L. Luo}
\affiliation{Beijing National Laboratory for Condensed Matter Physics, Institute of Physics, Chinese Academy of Sciences, Beijing 100080, P.R. China}
\author{A. Yazdani}
\affiliation{Joseph Henry Laboratories of Physics,  Department of Physics, Princeton
University, Princeton, NJ 08544, USA} 
\author{N.P. Ong}
\affiliation{Joseph Henry Laboratories of Physics,  Department of Physics, Princeton
University, Princeton, NJ 08544, USA} 
\author{N.L. Wang}
\affiliation{Beijing National Laboratory for Condensed Matter Physics, Institute of Physics, Chinese Academy of Sciences, Beijing 100080, P.R. China}
\author{M.Z. Hasan} 
\email [To whom correspondence should be addressed: ]  {mzhasan@Princeton.EDU}
\affiliation{Joseph Henry Laboratories of Physics,  Department of Physics, Princeton
University, Princeton, NJ 08544, USA} \affiliation{Princeton Center
for Complex Materials, Princeton University, Princeton, NJ 08544,
USA}


\date{14$^{th}$ August, 2008}


\begin{abstract}
We present a systematic photoemission study of the newly discovered high $T_c$ superconductor class
(Sr/Ba)$_{1-x}$K$_x$Fe$_2$As$_2$. By utilizing a unique photon energy range and scattering geometry we resolve the details of the single particle dynamics of interacting electrons on the central Fermi surfaces of this series which shows overall strong coupling behavior (2$\Delta_o$/$k_B$$T_c$ $\sim$ 6). Quasiparticle dispersion \textit{kinks} are observed in a binding energy range of 15 to 50 meV which matches the \textit{magnetic excitation} energy scales (parameterized by $J_1$,$J_2$). The size of the Cooper pair wavefunction is found to be less than $20$ $\AA$ indicating a short in-plane scale \textit{un}characteristic of a BCS-phonon scenario but suggestive of a phase factor in the global order parameter. The kink likely reflects contributions from the strongly frustrated fluctuating spin excitations and the soft phonons around 20-40 meV.


\end{abstract}

\maketitle

\begin{figure}[t]
\includegraphics[width=9.2cm]{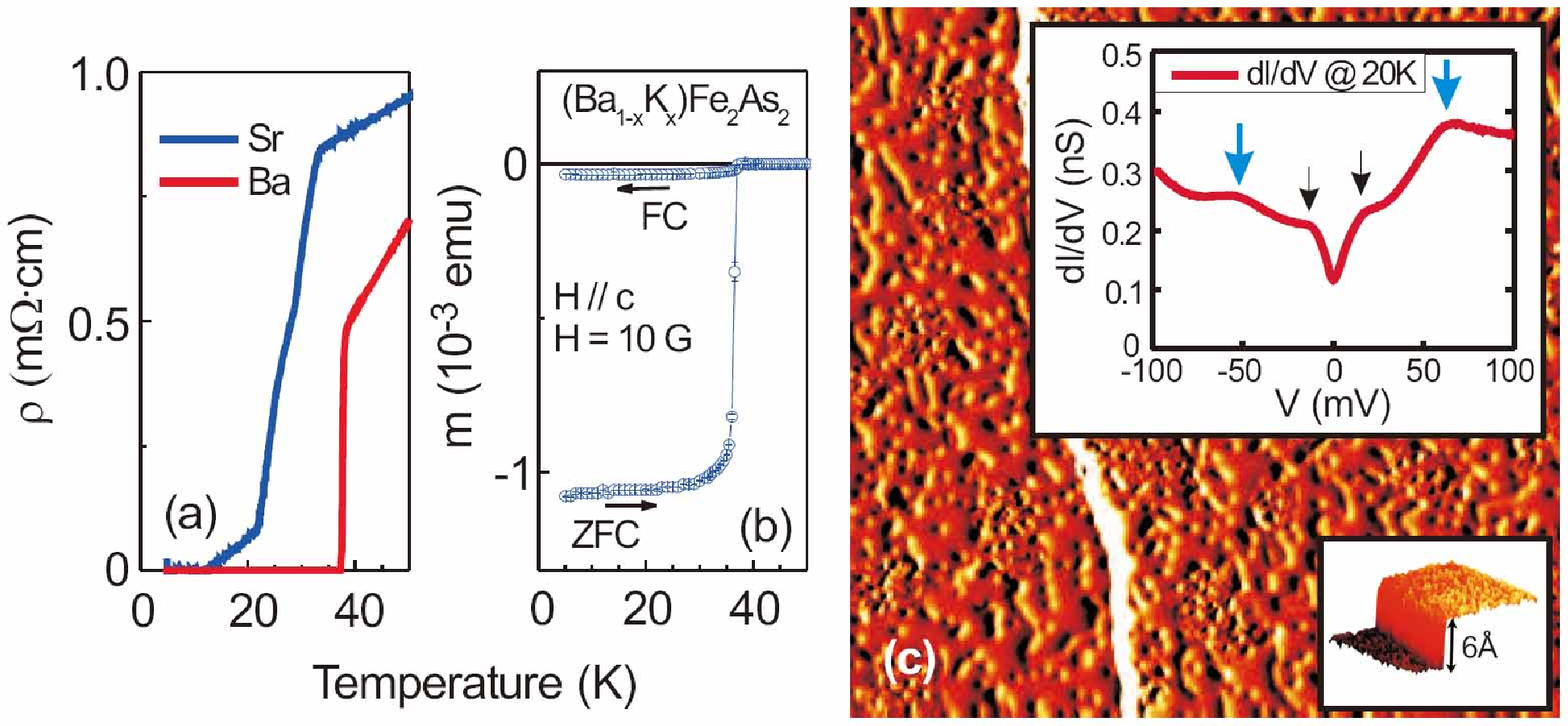}
\caption{\textbf{Phase transition, magnetization profiles and
surface quality}. (a-b) Bulk $T_c$ of crystalline $(Ba,K)Fe_2As_2$
and $(Sr,K)Fe_2As_2$ was determined based on the resistivity and
magnetization profiles. $(Ba,K)Fe_2As_2$ samples exhibited $T_c$ =
37K and $\delta$$T_c$ $\sim$ 1K whereas $(Sr,K)Fe_2As_2$ samples
exhibited a broad ($\sim$ 10K) transition with a $T_c$ $\sim$ 26K.
(c) Surface quality was studied by atomic-resolution STM
measurements which exhibited a high degree of flatness and confirmed
the suitability for spectroscopic measurements. The derivative of an
STM image is shown which was taken on a 500$\AA$$\times$500$\AA$
patch. The inset shows low-temperature electronic gap on the order
of 2$\Delta$$\thickapprox$30 meV in superconducting samples. Sample
batches with $\delta$$T_c$ $\sim$ 1K and smooth STM images were
selected for UHV cleaves in our ARPES studies.}
\end{figure}

Recent discovery of superconductivity ($T_c$ up to 55K) in
iron-based layered compounds promises a new route to high
temperature superconductivity \cite{kami, ren, chen, gfchen, cruz}. This
is quite remarkable in the view that the $T_c$ in the pnictides is
already larger than that observed in the single-layer cuprates.
These superconductors belong to a comprehensive class of materials
where many chemical substitutions are possible. Preliminary
studies suggest that the superconducting state in these materials competes with a magnetically ordered state, and the proper description of the magnetically ordered state lies somewhat in between a strong correlation mediated interacting local moment magnetism and quasi-itineracy with a high degree of stripe-like frustration \cite{cruz, singh,
kuroki, mazin, yao, eremin, bernevig, phonon_unlikely, ma, tesan, yildirim}.

Angle-resolved photoemission spectroscopy (ARPES) is a powerful tool
for investigating the microscopic electronic behavior of layered superconductors
\cite{arpes, xjz, ding, kaminski}. In this work we report
single-particle electronic structure results focusing on the details
of the low-lying quasiparticle dynamics and the pair gap
formation on very high quality ($\delta$$T_c$ $\lesssim$ 1K and surface-RMS $\lesssim$ 2$\AA$) single domain single crystal samples, which allow us to gain insight into connections between the superconductivity and magnetism not
addressed by other spectroscopic work \cite{xjz, ding, kaminski, lu}.
Besides a magnitude-oscillating superconducting gap in crossing from the zone center to the zone corner, we observe that the quasiparticles are strongly scattered by collective processes around 15 to 50 meV binding energy range depending on the Fermi surface sheet. Our high resolution and systematic quasiparticle data suggest that a Cooper pair in this superconductor is tightly bound ($\lesssim$ 4a$_o$), reflecting a non-phononic character of the underlying pairing potential in light of retardation or screening effects. Overall results can be self-consistently interpreted in a phase-shifting order parameter scenario.

\begin{figure*}[t]
\includegraphics[scale=0.8,clip=true, viewport=-0.5in 0.05in 14.6in 3.5in]{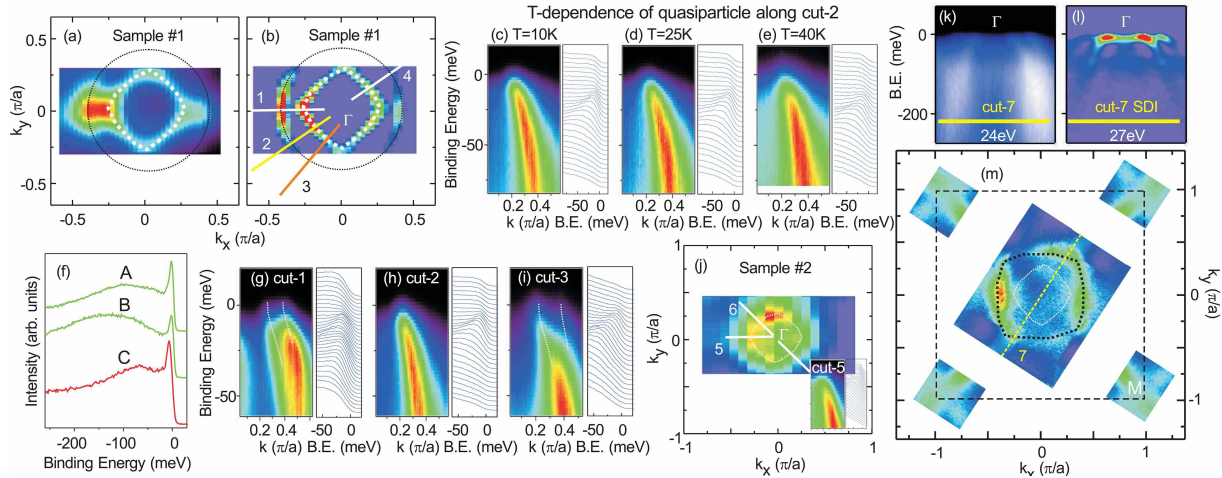}

\caption{\textbf{Quasiparticle behavior on the $\Gamma$-point Fermi
sheets in $(Sr/Ba,K)Fe_2As_2$}. (a) Momentum distribution of
quasiparticles within 15 meV of Fermi level in $(Ba,K)Fe_2As_2$. (b)
Second-derivative image approximation of the Fermi surface topology
around the $\Gamma$-point. (c-e) Quasiparticle dispersion along cut-2 and its temperature evolution. (f) High-resolution fine-step binding energy scans shown for some selected k-space points near the [1,0] and [1,1] axes (A, B) on the outer FS surrounding the $\Gamma$ point, and on a separate FS close to the M-point (C). (g-i) Quasiparticle intensity profiles along k-space cuts 1 to 3. The
k-space cut-2 strongly suppresses the outer FS and provides a clear spectroscopic look at the quasiparticle that forms the innermost FS. Because of the spectral clarity this quasiparticle can be studied in quantitative detail. (j) Fermi surface image taken on
$(Sr,K)Fe_2As_2$. (k,l) Wide k-range coarse-step scans are shown which was used for locating the Fermi crossings. (m) Electron distribution map, n(k), within 15 meV of Fermi energy over the complete Brillouin Zone.}
\end{figure*}

ARPES measurements were performed using 18 to 60 eV photons with
better than 8 to 15 meV energy resolution respectively and overall
angular resolution better than 1\% of the Brillouin zone. Most of the data were taken at the Advanced Light Source beamline 12.0.1 and a limited data set was taken at SSRL beamline 5-4 for cross-checking, using a Scienta analyzer
with chamber pressures lower than 5x10$^{-11}$ torr. Linearly polarized photons were used for all the study. The angle between the $\overrightarrow{E}$-field of the incident
light and the normal direction of the cleaved surface was set to
about 45 degrees (at 12.0.1). Single crystalline samples of
Ba$_{1-x}$K$_x$Fe$_2$As$_2$ ($T_c$=37K) and
Sr$_{1-x}$K$_x$Fe$_2$As$_2$ ($T_c$=26K) were used for this systematic study.
Cleaving the samples in situ at 15K resulted in shiny flat surfaces.
Cleavage properties were characterized by atomic resolution STM
measurements and the surface was found to be flat with an RMS
deviation of 1\AA (Fig.1(c)). Rarely observed steps of size 6$\AA$
were seen on the otherwise flat surface. The utilization of a unique
scattering geometry and specific photon energy range allowed us to
suppress one of the FS sheets so that the other can be studied in full detail.

\begin{figure*}[t]
\includegraphics[width=18.2cm]{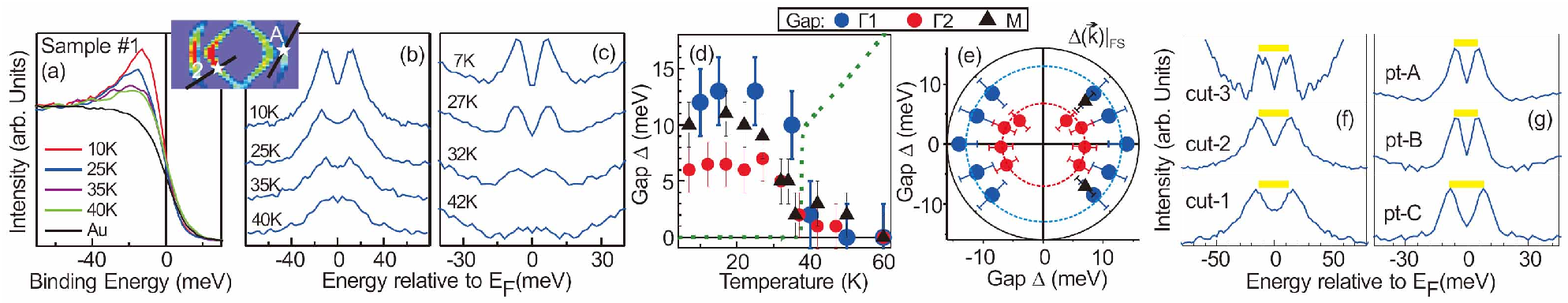}
\caption{\label{fig:Sb_Fig2} \textbf{Multi-gap pairing structure:} (a) Temperature dependence of quasiparticles (cut-2) near the Fermi level through the
superconducting transition. Below $T_c$ samples exhibit coherence-peak-like behavior similar to what is observed in some cuprates. Temperature dependences of the gap at the k-space location of the inner-most ($\Gamma1$) FS (b) and the outer central ($\Gamma2$) FS (c) are estimated by symmetrized spectral weight around the Fermi level. (d) The temperature dependences of the gaps measured at different FS locations ($\Gamma1$-FS, blue; $\Gamma2$-FS, red; and near-$M$, black) are plotted along with the bulk resistivity curve (green). A fluctuation regime above $T_c$ is observed. (e) The azimuthal k-dependences of gaps, $\Delta(k)$, are shown for different FS sheet locations ($\Gamma1$, $\Gamma2$, $M$). Selected EDCs are shown for the k-dependence of gaps on the $\Gamma1$ (f) and outer Fermi surfaces (g). The cuts and pts refer to Fig.2.}
\end{figure*}

Quasiparticle behavior around the $\Gamma$-FS sheets is shown in Figure 2.
Two square-like FS sheets were clearly resolved near the center of the BZ. An azimuthal
variation of ARPES intensity around the FS pockets was observed.
This variation is most pronounced while the data are taken at the
particular photon-electron scattering geometry described above. A
comparison of quasiparticle dispersion measured along the various
$\overrightarrow{k}$-space cuts suggests that roughly along a cut 45-degrees to the
$\Gamma$ to $(\pi, 0)$-line provides a clear spectroscopic look at
the quasiparticle dispersion and lineshape behavior on the inner-FS.
A bend in the dispersion (E vs. $\overrightarrow{k}$) could be observed in the data
which is not resolved in cut-1 or 3 due to the spectral overlap with
the outer-FS. The measured quasiparticles in Sr$_{1-x}$K$_x$Fe$_2$As$_2$
were found to be broad and no bend could be resolved.

The temperature evolution of low-lying quasiparticles through the superconducting transition is shown in Figure 3. Quasiparticles gain spectral
weight upon cooling below the transition temperature ($\sim$ 37K).
The low-temperature quasiparticle peak width is about 10 meV. The
opening of the superconducting gap is best viewed upon
symmetrization \cite{norman} of the near-$E_F$ data and a gap magnitude of about
12$\pm$2 meV is quite evident at low temperatures. This value is consistent with the average gap ($<$ 15 meV, see Fig.1(c)) we observe with STM on the same batch of samples. The observed gap value is found to be largest in the inner-most central FS ($\Gamma1$ band), then decreases on the next FS ($\Gamma2$ band) moving outward toward the M-point, and then increases again on the corner FS location. This oscillating gap structure is consistent with (but not a unique fit to) an order-parameter that takes the in-plane form of $\Delta_o$$cos(k_x)cos(k_y)$. The reduction of the gap value on the outer central FS is consistent with a cos$\times$cos form. We caution that the ARPES data do not rule out the possibility of an out-of-plane ($k_z$) node in the order parameter. Existence of such a node may explain the in-gap $T^3$ behavior of NMR data \cite{nmr}.


A closer look at the quasiparticle dispersion behavior is presented in Figure 4. A bend in dispersion is evident in the momentum distribution curves (MDC) taken on a crossing near the $\Gamma1$-FS (cut-4). Each MDC could be fitted with a single Lorentzian over a wide binding energy range and, as in the raw data sets, the fitted peak positions trace a kink around 40$\pm$15 meV. This is further confirmed by examining the peak position of the real part of the self-energy which is also observed to be around 40$\pm$15 meV. Although it is less clear, the MDC width plotted as a function of the electron binding energy is found to exhibit a drop below 30 meV that is consistent with a kink at higher energy seen in the raw data. At temperatures above $T_c$ the kink shifts to somewhat lower energies. As the temperature is raised further the MDCs are broadened making its identification or analytic extraction from our experimental data difficult and unreliable. In the MDC widths (Fig.4(g)) an increase is observed at very low energies which was found to be unrelated to the existence of the kink but rather related to some residual signal from the tail of the
quasiparticles on the outer FS. The STM data in Fig.1(c) also exhibit a satellite structure around 40-50 meV loss-energy range (with respect to the quasiparticle peak position) roughly consistent with the observed ARPES kink. Assuming that the kink reflects coupling to some bosonic-like modes one can estimate the coupling strength: $\lambda'_{eff}$$\gtrsim$ (0.7/0.45 - 1)$\sim$ 0.6. This coupling is about a factor of two to three larger than the electron-phonon coupling ($\lambda_{ph}$$\sim$ 0.2) calculated for the Fe-As phonons near 20-40 meV \cite{phonon_unlikely}. A careful look at the outer central FS ($\Gamma2$ band, cut-8) also reveals a kink around 18 $\pm$ 5 meV. This kink is revealed when the band associated with the inner-FS sheet is suppressed by a choice of incident photon energy such as in the data taken near 18 eV. The observed kink energies seems to scale (40 meV and 18 meV) with the superconducting gap energies (12 meV and 6 meV) on the two central Fermi surfaces (Fig.4).

\begin{figure*}
\includegraphics[scale=0.88,clip=true, viewport=0in 0.04in 14.9in 2.5in]{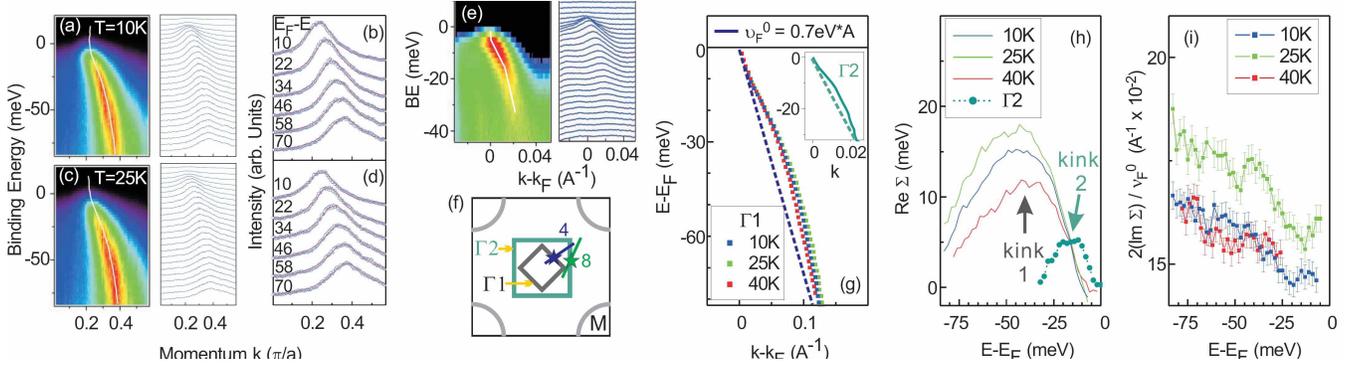}
\caption{\label{fig:Sb_Fig3} \textbf{Dispersion kinks}. (a),(c) Quasiparticle band dispersion along cut-4 (k-space cut-4 is approximately along the $Q_{AF}$-vector of the undoped compound) and the corresponding MDCs at 10K and 25K. All the MDCs can be fitted by a single Lorentzian with linear background. (b),(d) The quasiparticle
lineshapes are presented at selected energies (10-70 meV). (e) Band dispersion for the $\Gamma2$ quasiparticle along cut-8 shows a small kink near 18meV.  (f) Cuts 4 and 8 are labeled on a FS schematic. (g) By tracing the peak positions, quasiparticle band dispersion is plotted at T=10K, 25K and 40K. At all temperatures, the dispersion curve
shows a "kink"-like feature at 40$\pm$10 meV. The gray dashed line
illustrates the "bare" band used for extracting the real part of the
quasiparticle self-energy. The smaller kink at 18$\pm$5meV on the $\Gamma2$ band is shown in the inset. (h) The real part of self energy is obtained by subtracting the gray dashed line from the experimental band dispersion. Peak position is used to define the "kink" position. (i) MDC width as a function of binding energy for quasiparticles on $\Gamma1$.}
\end{figure*}

A strong-coupling kink phenomenology is observed in the electron dynamics of high $T_c$ cuprates which occurs around 60$\pm$20 meV and is often attributed to phonons or magnetism or polarons with $\lambda'_{eff}$$\sim$ 1 to 1.5 \cite{arpes}. In cuprates the superexchange coupling is on the order of 130 meV, whereas the optical phonons are in the range of 40 to 80 meV overlapping with the kink. In the pnictides, although a $T_c$ value of 37K is not outside the phonon-induced strong-coupling pairing regime, the vibrational modes of the the FeAs plane are rather soft ($\leq$ 35 meV) making electron-phonon interaction \cite{phonon_unlikely} an unlikely source of the majority part of the quasiparticle's self-energy beyond 40 meV, considering the observed coupling $\lambda'_{eff}$$\gtrsim$0.6 for the FeAs compounds here. The parent compounds of superconducting FeAs exhibit a robust SDW groundstate \cite{cruz} due to a $\overrightarrow{Q}$=($\pi$, $\pi$) inter-band instability or due to the interaction of quasilocalized moments. The SDW short range order seems to survive well into the superconducting doping regime \cite{phd}. The doping evolution of the Fermi surface lacks robust nesting conditions for purely band-magnetism to be operative at these \textit{high dopings} and the relevant magnetism likely comes from the local exchange energy scales in a doping induced frustrated background. Therefore, quite naturally, strong spin fluctuations in the presence of electron-electron interaction are important contributors to the electron's self-energy and a natural candidate for the constituent of the kinks. In accounting for the parent SDW groundstates of these materials the known values of $J_1$ and $J_2$ are on the order of 20 to 50 meV \cite{ma, yildirim}. This is the energy range where we have observed the quasiparticle kinks (Fig.4). In an itinerant picture, there exists a Stoner continuum whose energy scales are parameterized by $J_1$ and $J_2$ whereas in a local picture, $J_1$ and $J_2$ reflect Fe-Fe and Fe-As-Fe superexchange paths and the groundstate is a highly frustrated doped Heisenberg magnet \cite{ma}. The proper description of the experimentally observed magnetism in these systems lies somewhere in between. In the photoemission process removal of an electron from the crystal excites the modes the electron is coupled to, so the observed quasiparticle breaks the locally frustrated magnetic bonds associated with $J_1$ and $J_2$ which then contributes a characteristic energy scale in the electron's self-energy. Since these characteristic magnetic scales are quite large ($\sim$ 400K) it is expected that the kink behavior in the electron's dispersion relation would survive above $T_c$ consistent with our experimental observation. Despite the high signal-to-noise quality of our data, it is premature to draw an intimate connection between the kinks and the resonant spin mode ($\sim$ 15 meV) observed in neutron scattering \cite{spinmode}.

High-resolution (Fig.-4) dispersion measurements further allow us to estimate the Fermi velocity of the normal state which is about 0.7 eV$\cdot\AA$. Using the observed superconducting gap (ARPES or STM data in Fig.1) we can estimate the average size of the Cooper pair wavefunction : $\xi= \frac{\hbar v_F}{\pi\Delta}$ by invoking the uncertainty relation \cite{tinkham}. Taking $v_F$(Fig.4) $\sim$ 0.7$\pm$0.1 eV$\cdot\AA$ and a gap (Fig.3) value of $\Delta$ $\sim$ 12$\pm$2 meV, this gives $\xi$ $\lesssim$ 20 $\AA$. This value is remarkably consistent with the high magnitude of H$c_2$ ($\sim$70T) \cite{hc2} reported in these same materials. The ARPES based Cooper pair scale and unusually high H$c_2$ clearly suggest that the Cooper pairs in this class of FeAs superconductors are tightly bound which is in contrast to the point-contact Andreev spectroscopy results on Sm-based FeAs superconductors exhibiting a conventional BCS ratio \cite{jhu}. The agreement between ARPES, bulk H$c_2$ and the bulk resistivity profile (Fig.3(d)) provides further support for our identification of the superconducting gap and its bulk-representative value through a surface-sensitive measurement such as ARPES. This also confirms that the ARPES observed gaps in superconducting materials \cite{xjz, ding, kaminski} are not the SDW gaps as theoretically claimed by some authors. More importantly, such a small Cooper pair size scale ($\sim$ 4$a_o$) is not known in any phonon-based BCS superconductor \cite{kivelson} but has only been observed in unconventional correlated superconductors. Our observed value is much smaller than that in the multi-band s-wave BCS-phonon superconductors such as MgB$_2$ \cite{kivelson}. In fact a combination of small Cooper pair size and in-plane nodeless superconductivity is consistent with an unconventional $\Delta_o$$cos(k_x)cos(k_y)$(in the unfolded BZ with one iron atom per unit cell)-type or s$_{x^2y^2}$ or s$_{\pm}$ wave states \cite{kuroki, mazin, yao, eremin, bernevig, tesan} since such an order parameter has a nearest-neighbor structure in real space and thus a reduction of the Coulomb interaction within the pair is naturally possible, so the electrons can come closer to each other leading to a short coherence scale. In cuprates pairing electrons come close to each other, and the short coherence length is achieved by introducing a node in the order parameter (\textit{d}-wave) leading to a reduction of Coulomb interaction within the pair. This is often the only choice in a single band system such as the cuprates or the organics. In pnictides, multiband structure can accommodate a phase change without the need for introducing a "node" \cite{coscosnode} on the Fermi surface, therefore an isotropic gap and short pairing scale can co-exist with a phase shifted order-parameter structure.

In summary, we have presented a single-particle study of high $T_c$
superconductor class (Sr/Ba)$_{1-x}$K$_x$Fe$_2$As$_2$. Our systematic ARPES data suggest an unusually small dimension of the Cooper pair, complex kink phenomena and an oscillating gap function all of which collectively point to an unconventional pairing potential. We have presented arguments that in the presence of Coulomb interaction and magnetism, the observed short pairing scale and a nearly-isotropic in-plane gap can be self-consistently realized if the order parameter contains a phase factor. Our results thus provide important clues to a novel route to high temperature superconductivity.


\newpage

\begin{figure*}
\includegraphics[width=14cm]{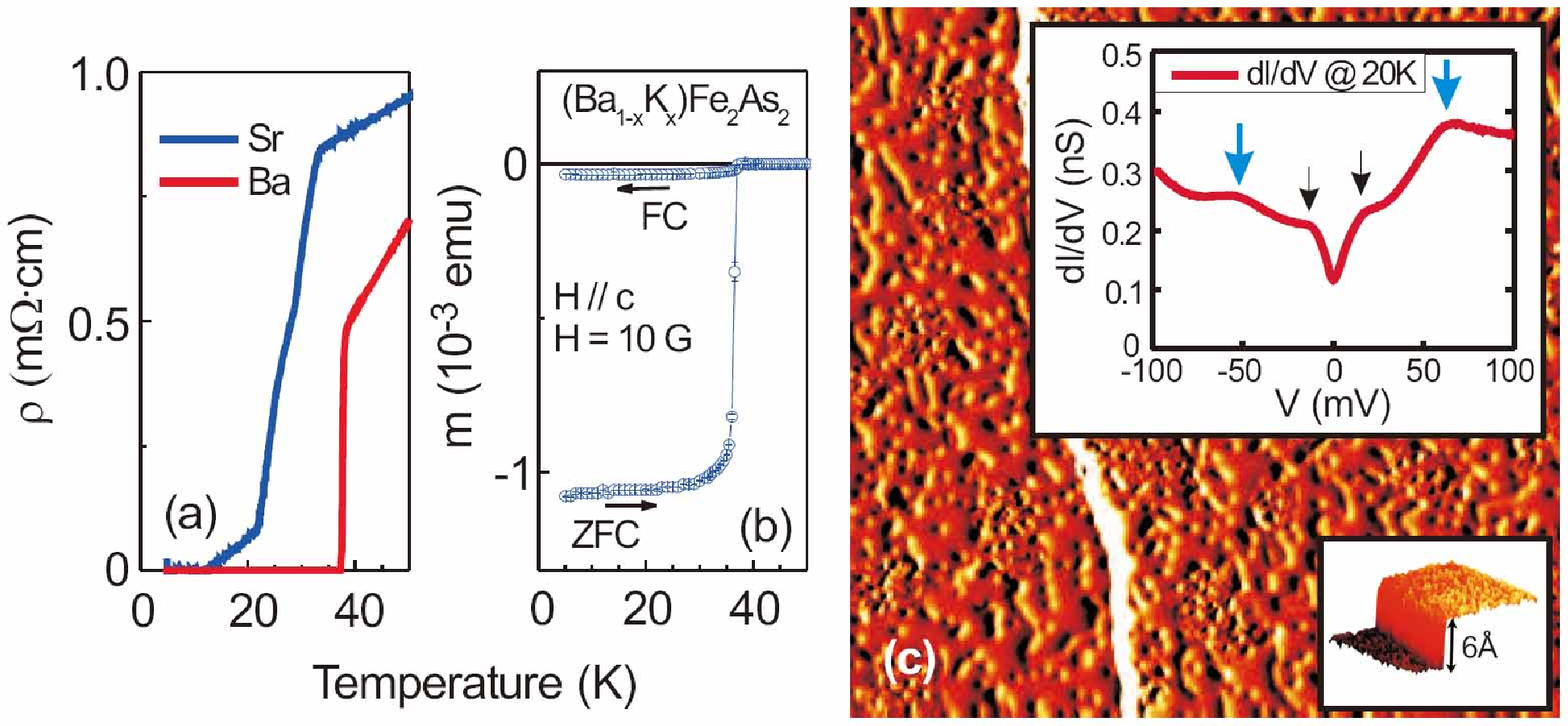}
\caption{\textbf{Enlarged View of Figure 1}. }
\end{figure*}

\begin{figure*}[t]
\includegraphics[scale=0.9,clip=true, viewport=-0.5in 0.05in 14.6in 3.5in]{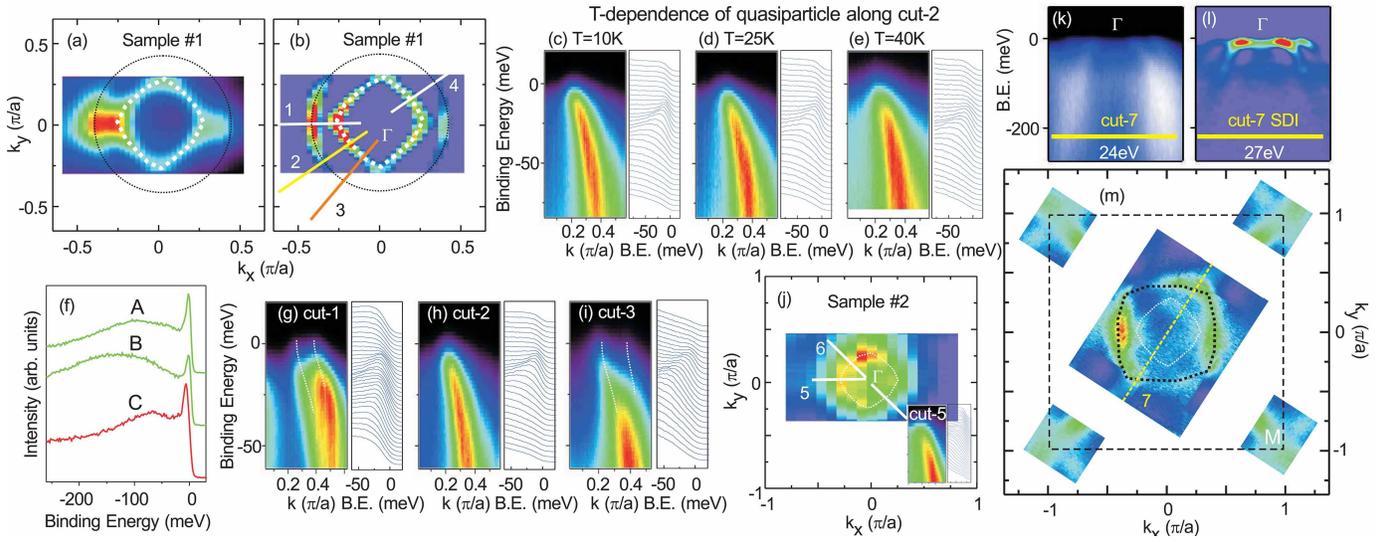}
\caption{\textbf{Enlarged View of Figure 2}. }
\end{figure*}

\begin{figure*}[t]
\includegraphics[width=19cm]{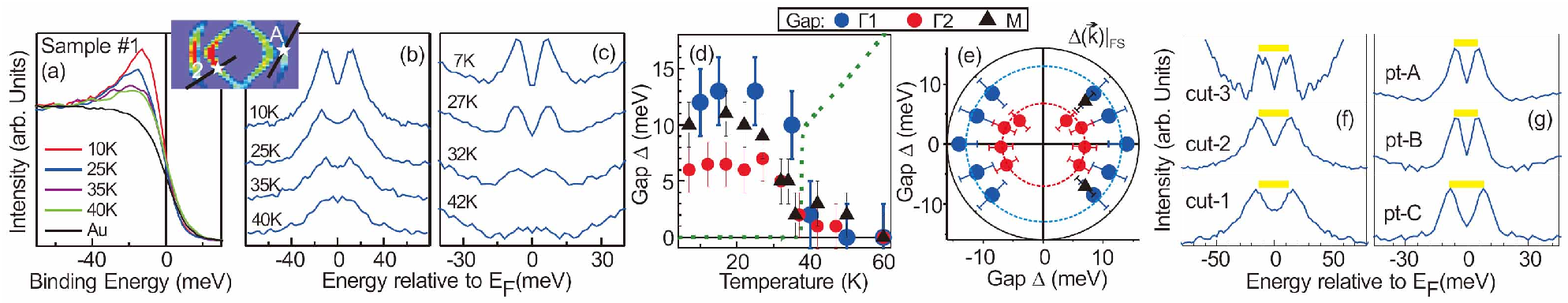}
\caption{\label{fig:Sb_Fig2} \textbf{Enlarged View of Figure 3} }
\end{figure*}

\begin{figure*}
\includegraphics[scale=0.92,clip=true, viewport=0in 0.04in 14.9in 2.5in]{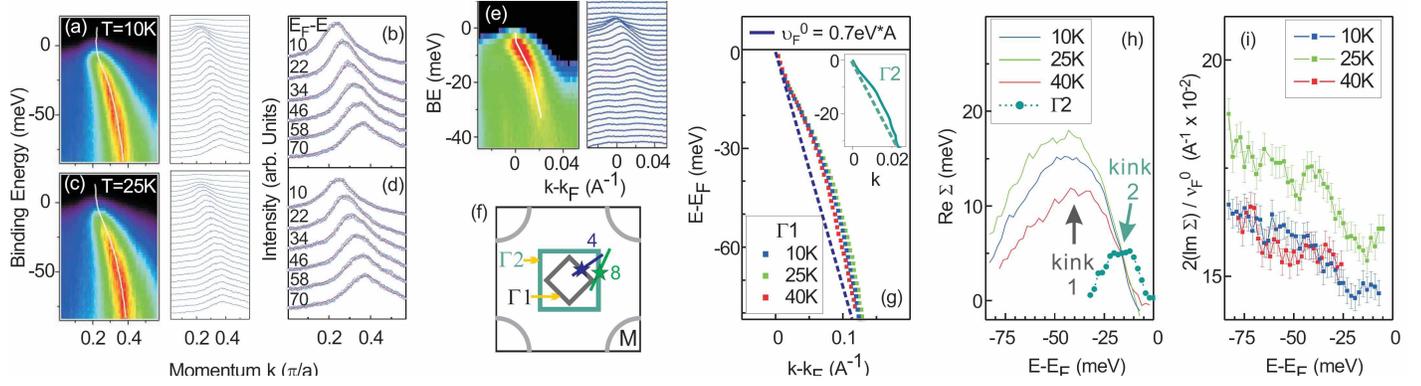}
\caption{\label{fig:Sb_Fig3} \textbf{Enlarged View of Figure 4}. }
\end{figure*}



\begin{thebibliography}{h}
\bibitem{coscosnode} The cos$\times$cos or s$_{\pm}$ gap does have a node in it but the node does not intersect the Fermi surface as we measure here (Figure 3). However, it is clear that nearest or next nearest neighbour structure of the gap lowers the kinetic energy as opposed to a constant on-site s-wave gap.
\end{thebibliography}
\end{document}